\documentclass[sigconf,nonacm]{acmart}

\usepackage[ruled,vlined,linesnumbered]{algorithm2e}
\usepackage{graphicx}
\usepackage{xspace}
\usepackage{multirow}
\usepackage[normalem]{ulem}
\usepackage{contour}

\contourlength{0.2ex}

\newcommand{\ulc}[2]{%
  {\color{#2}\uline{\phantom{#1}}}%
  \llap{\contour{white}{#1}}%
}

\newcommand{\added}[1]{\textcolor{black}{#1}}

\definecolor{colorBlue}{HTML}{1f77b4}
\definecolor{colorOrange}{HTML}{ff7f0e}
\definecolor{colorGreen}{HTML}{2ca02c}
\definecolor{colorRed}{HTML}{d62728}
\definecolor{colorBright}{HTML}{fde725}
\definecolor{colorMedium}{HTML}{21918c}
\definecolor{colorDark}{HTML}{440154}

\newtheorem*{remark}{Remark}

\renewcommand{\epsilon}{\varepsilon}
\newcommand{\Scoref}{\ensuremath{\varsigma}\xspace}
\newcommand{\Let}[2]{#1 $\leftarrow$ #2}
\newcommand{\algo}[1]{\textsc{#1}}
\newcommand{\Gaussian}{\textsc{GaussianGen}\xspace}
\newcommand{\HephGrad}{\textsc{Hephaestus-Gradient}\xspace}
\newcommand{\HephAnn}{\textsc{Hephaestus-Annealing}\xspace}

\newcommand{\query}{\ensuremath{\vec{q}}\xspace}
\newcommand{\point}{\ensuremath{\vec{x}}\xspace}
\newcommand{\Expansion}{\ensuremath{\operatorname{Expansion}}\xspace}
\newcommand{\RC}{\ensuremath{RC}\xspace}
\newcommand{\LID}{\ensuremath{LID}\xspace}
\newcommand{\ehard}{\ensuremath{\epsilon\text{-hardness}}}
\newcommand{\dataset}[1]{\texttt{#1}}

\newcommand{\hnsw}{\textsc{HNSW}\xspace}
\newcommand{\ivf}{\textsc{IVF}\xspace}
\newcommand{\messi}{\textsc{MESSI}\xspace}
\newcommand{\dstree}{\textsc{DSTree}\xspace}

\AtBeginDocument{%
  }

\copyrightyear{2025}
\acmYear{2025}
\setcopyright{cc}
\setcctype{by}
\acmConference[KDD '25]{Proceedings of the 31st ACM SIGKDD Conference on Knowledge Discovery and Data Mining V.2}{August 3--7, 2025}{Toronto, ON, Canada}
\acmBooktitle{Proceedings of the 31st ACM SIGKDD Conference on Knowledge Discovery and Data Mining V.2 (KDD '25), August 3--7, 2025, Toronto, ON, Canada}
\acmDOI{10.1145/3711896.3737383}
\acmISBN{979-8-4007-1454-2/2025/08}

\begin{document}

\title{Evaluating and Generating Query Workloads \\ for High Dimensional Vector Similarity Search}

\author{Matteo Ceccarello}
\email{matteo.ceccarello@unipd.it}
\orcid{0000-0003-2783-0218}
\affiliation{%
	\institution{University of Padova}
	\city{Padova}
	\country{Italy}
}

\author{Alexandra Levchenko}
\email{alexandra.levchenko@isep.fr}
\orcid{0000-0002-4230-338X}
\affiliation{
    \institution{Isep, LISITE}
    \city{Issy-Les-Moulineux}
    \country{France}
}

\author{Ioana Ileana}
\email{ioana.ileana@parisdescartes.fr}
\orcid{XXX}
\affiliation{
    \institution{Universit\'e Paris Cit\'e}
    \city{Paris}
    \country{France}
}

\author{Themis Palpanas}
\email{themis@mi.parisdescartes.fr}
\orcid{0000-0002-8031-0265}
\affiliation{
    \institution{Universit\'e Paris Cit\'e}
    \city{Paris}
    \country{France}
}

\renewcommand{\shortauthors}{Matteo Ceccarello, Alexandra Levchenko, Ioana Ileana, and Themis Palpanas}

\begin{abstract}
Similarity search lies at the heart of many modern applications,
ranging from databases to deep learning to data series analysis.
As such, a vast effort has been invested in developing algorithms,
data structures and implementations to speed up this crucial subroutine.
To empirically validate these approaches, several benchmarking efforts
have been initiated covering a wide array of datasets.
In this paper, we observe that usually little control is exercised on the hardness of the workloads with which methods are tested and compared.
To address this issue, we first evaluate several query hardness 
measures with respect to their ability to capture the empirical 
hardness of a query, i.e. the effort invested by an index data structure to provide an answer.
Then, we propose two methods, deemed \HephAnn and \HephGrad, for synthesizing query workloads so that they meet a user-specified hardness target.
Both methods allow to produce workloads with the desired hardness:
we find that \HephGrad is faster, while \HephAnn
makes fewer assumptions on the target hardness measure.
The resulting workloads can be used to gain insights into the behavior 
of similarity search algorithms.
This paper appeared in the proceedings of KDD 2025~\cite{DBLP:conf/kdd/CeccarelloLIP25}.
\end{abstract}



\maketitle

\section{Introduction}

High-dimensionality vector similarity search is a fundamental task in a wide range of applications, from
information retrieval and recommendation systems to computational biology and
computer vision~\cite{DBLP:journals/dagstuhl-reports/BagnallCPZ19,DBLP:journals/sigmod/PalpanasB19,DBLP:conf/wims/EchihabiZP20}. 
The goal of similarity search is to identify items in a dataset
that are most similar to a given query, based on some defined similarity
measure. Given the prominence of this task, a rich ecosystem of algorithms,
index data structures, and implementations has flourished in recent years~\cite{DBLP:journals/pvldb/EchihabiZPB18,DBLP:journals/pvldb/EchihabiZPB19,isaxfamily,DBLP:journals/debu/00070P023}.

Alongside the development of index data structures and algorithms, the necessity
of establishing a common testbed to evaluate different implementations has
spurred a variety of benchmarking efforts~\cite{DBLP:journals/is/AumullerBF20,
DBLP:journals/tkde/LiZSWLZL20,
DBLP:journals/pvldb/EchihabiZPB18,DBLP:journals/pvldb/EchihabiZPB19,DBLP:conf/nips/SimhadriWADBBCH21}.
These benchmarks are now in widespread use and provide the community with a very
useful resource to assess the merits of different algorithms and implementations
in a variety of scenarios, encompassing
different datasets and workloads.

The goal of a benchmark is to find, for a given workload, what is the performance
of different implementations under different parameter settings.
Performance can be measured in terms of time to run queries, or of number of executed distance
computations, or again in terms of the quality of the answer itself, for approximate index data structures.
This allows to study the behavior of algorithms under different circumstances, as well as the tradeoffs involved in configuring the implementations,
while allowing the community to find the fastest implementation on a given workload.

In this paper we focus on studying the \emph{workloads}, i.e. the set of
queries, that are used to evaluate competing algorithms. In fact, the common
practice to prepare a workload follows the tradition set by machine learning: a
dataset is partitioned randomly in two parts, a larger one to be indexed and
smaller one to be used as queries.
This approach is sensible in that it gives a set of readily available queries
that come from the same distribution of the data.
However, how hard are these queries, both intrinsically and for a given index?

As we shall see with our experiments,
these workloads do not exercise
the full spectrum of possible behaviors of the index data structures,
and real-world workloads have been found to be considerably harder\added{~\cite{amazonhard,VIBE,DBLP:journals/debu/0001C23,DBLP:journals/debu/BruchNRV24}.}
Furthermore, these workloads do not provide focused information about queries with a
specific hardness.
Finally, the same query point can be hard
or easy depending on the value of $k$.
These considerations are relevant in order to stress-test index structures, better understand their behavior, and to design and develop better-performing versions.

The culprit is that the workload selection procedure used so far does not allow
to exercise any control on the hardness of the queries. We thus 
propose a way to \emph{generate} workloads with a pre-specified level of
hardness. 
The first step is measuring the hardness of a workload: \emph{empirical}
hardness measures evaluate the work performed by actual index structures,
\emph{explicative} measures on the other hand try to capture geometric
properties relating queries and data.
While \emph{empirical} measures are what a user is ultimately interested in,
\emph{explicative} measures allow to further interpret the results:
why are some queries doing more work than others?
What are the characteristics
that make them easy or hard to answer?

Using these measures, we propose methods to generate query workloads
both in a index-agnostic and a index-aware way.
In the former case, queries should present intrinsic characteristics that 
make them easy or hard: different index data structures might then be
able to deal with them with different degrees of efficiency.
In the latter case, queries should be hard (or easy) for a \emph{specific}
index data structure: this allows to identify bottlenecks and weak
spots during index development and tuning.

Our contributions are articulated as follows:
\textbf{(a)} We provide an overview of different measures to assess the hardness of queries, both in an index-independent and index-centered way (Section~\ref{sec:preliminaries});
\textbf{(b)} We propose methods to generate synthetic queries of given target hardness (Section~\ref{sec:workloads-generators}), \added{providing an easy to
use Python implementation\footnote{\url{https://github.com/cecca/hephaestus/}}};
\textbf{(c)} We experimentally evaluate the relation between hardness measures and actual query difficulty, and test the effectiveness of the methods we propose in generating query workloads (Section~\ref{sec:experiments}).

\section{Related Work}

High-dimensional vector nearest neighbor search is a crucial subroutine in many different contexts, with a wide array of different approaches being proposed. 
For a recent account of the latest developments see~\cite{isaxfamily,DBLP:journals/debu/0001C23,DBLP:journals/debu/00070P023} and references therein.

Among the many approaches devised for \emph{approximate} nearest neighbor search, we focus on the ones implemented in the \textsc{Faiss} library~\cite{DBLP:journals/corr/abs-2401-08281}, including Hierarchical Navigable Small World (HNSW) graphs~\cite{DBLP:journals/pami/MalkovY20}
and an Inverted File~\cite{DBLP:conf/iccv/SivicZ03,DBLP:journals/pami/JegouDS11}.
The \textsc{Faiss} library has been adopted as a baseline by some recent benchmarking efforts~\cite{DBLP:conf/nips/SimhadriWADBBCH21}.

Another line of work~\cite{DBLP:journals/pvldb/WangWPWH13,DBLP:conf/bigdataconf/PengFP18,DBLP:journals/tkde/PengFP21, DBLP:conf/icde/PengFP20} focused on building data structures providing \emph{exact} answers to queries.
Here we focus on \messi~\cite{DBLP:journals/vldb/PengFP21},
which builds a tree-based index on iSAX words~\cite{DBLP:conf/kdd/ShiehK08},
and on \dstree~\cite{DBLP:journals/pvldb/WangWPWH13}.

A popular benchmarking effort is ann-benchmarks~\cite{DBLP:journals/is/AumullerBF20}, which provides a collection of datasets and query workloads.
Since its inception, over 40 algorithms have been included in the benchmark.
We will use some query workloads of ann-benchmarks as a baseline in our experimental evaluation.
A similar benchmarking effort is described in~\cite{DBLP:journals/tkde/LiZSWLZL20}, focusing on the Euclidean distance case.
Other benchmarking efforts are~\cite{DBLP:journals/pvldb/EchihabiZPB18,DBLP:journals/pvldb/EchihabiZPB19},
evaluating both approximate and exact approaches on a variety of datasets, which we also include in this paper.
Recently, big-ann-benchmarks~\cite{DBLP:conf/nips/SimhadriWADBBCH21} scaled the benchmarking effort to
billion-scale data, extending the range of tasks to include filtering further the results based on
categorical features, out-of-distribution queries, and dynamic data updates.
VIBE~\cite{VIBE} focuses on datasets derived from embedding models characteristic of modern applications.

Dimensionality measures such as
the Local Intrinsic Dimensionality~\cite{DBLP:conf/icdm/Houle13},
the Relative Contrast~\cite{DBLP:conf/icml/HeKC12},
the query Expansion~\cite{DBLP:conf/soda/AhleAP17,DBLP:journals/is/AumullerC21},
and the $\epsilon$-hardness~\cite{DBLP:journals/vldb/ZoumpatianosLIP18}
capture features of the distribution
of distances from a query that relate
to how hard it is to discern nearest neighbors from the rest
of the points.
\added {The \emph{Steiner}-hardness~\cite{DBLP:journals/pvldb/WangWCWPW24} is specifically designed to investigate the performance of graph-based indices.}
The relation between these measures and the actual performance
of indices was investigated in~\cite{DBLP:conf/sisap/0001C19,DBLP:journals/is/AumullerC21}: in the present paper we build upon and expand their conclusions.

The goal of generating query workloads has also been pursued in Zoumpatianos et al.~\cite{DBLP:journals/vldb/ZoumpatianosLIP18}:
given a query, the dataset is modified so that the query achieves the desired hardness.
\added{
However, modifying the dataset may not be desirable.
Moreover, this method incurs significant runtime costs.
}
In the present paper, we take the opposite approach: while keeping the dataset fixed, we carefully place queries in the metric space so to achieve the desired hardness level.
Given the fundamentally different scenario that we consider, 
we will not compare directly with~\cite{DBLP:journals/vldb/ZoumpatianosLIP18}.

\section{Preliminaries}\label{sec:preliminaries}

Let $(\mathcal{X}, d)$ be a metric space, with $d$ being the distance function.
In this paper, in particular, 
we focus on the $D$-dimensional Euclidean space $\mathbb{R}^D$ under the Euclidean distance,
and on $S^{D-1}$, the unit sphere in $D$ dimensions under the angular distance.

We denote with $S \subset \mathcal{X}$ a dataset,
and with $\point \in \mathcal{X}$ we denote a point in the metric space.
Note that in different communities \point is referred to with different names: it can be a point, a vector, or a data series if the order of coordinates is relevant.
Given that in this paper we focus on the Euclidean and angular distances, for which the order of the coordinates is not relevant, we will refer to elements of $\mathcal{X}$ as \emph{points}.

For a dataset $S$, a query point $\query \in \mathcal{X}$ and a parameter $k$,
the $k$-nearest neighbor problem entails finding the $k$ points in $S$ that are
closest to $\query$ according to the distance function $d$, with ties broken arbitrarily.
We denote with $r_1, r_2, \dots, r_k$ the distance between \query{} and its first, second, \dots, $k$-th nearest neighbor.

Such queries can be answered exactly in time $O(n)$ by simply evaluating the
distance between $\vec{q}$ and all points $\vec{x} \in S$. However, for large
datasets it is often desirable to have sublinear query time, often allowing some approximation in the answers. 

In the case of approximate approaches, the quality of the answers is usually
measured using the \emph{recall}, defined as
$ \frac{|\{ 1 \le i \le k : r_i' \le r_k\}|}{k} $.
In other words, the recall is the fraction of
returned points whose distance from the query is $\le r_k$, the distance of the
true $k$-th nearest neighbor.

\subsection{Hardness Measures}\label{sec:hardness-measures}

We now survey several established \emph{explicative} hardness measures and introduce the \emph{empirical hardness}.
We deem \emph{explicative} those measures that are based only on intrinsic properties of a query in relation to a dataset, like the distribution of distances. Such measures, which are independent of any index data structure, can help to \emph{explain} why a given query can be expected to be hard or easy.
On the other hand, \emph{empirical} measures reflect the actual performance
of a given index data structure on a given query.
In the following sections we will then study the relationship between these measures.

\subsubsection{Local intrinsic dimensionality}

For a given query point $\query\in \mathcal{X}$ we consider the distribution of distances to \query
within the metric space, where the distribution arises by sampling $n$ points
from the metric space  under a certain probability distribution.
Let $F: \mathbb{R} \to [0,1]$ be
the cumulative distribution function of distances to $\query$.

\begin{definition}[\cite{DBLP:conf/icdm/Houle13}]
	The local intrinsic dimensionality of $F$ at distance $r$
	is
	\[
		ID_F(r) = \lim_{\epsilon \to 0} \frac{
			\ln\left(F((1+\epsilon)r) / F(r) \right)
		}{
			\ln\left((1+\epsilon)r) / r \right)
		}
	\]
	whenever the limit exists.
\end{definition}

Intuitively, the above measure is related to how quickly the probability mass
(i.e. the fraction of points that are within the ball of radius $r$ and the
ball of radius $(1+\epsilon)r$ around the query point) increases around the
query point.

A large LID value means that it is hard to distinguish points at distance
$r$ from the query from the rest of the dataset. Therefore, a hard query is expected to have a large LID value.

The LID can be estimated with a Maximum Likelihood
Estimator~\cite{DBLP:conf/nips/LevinaB04,DBLP:journals/datamine/AmsalegCFGHKN18}.
Let $r_1\le \dots \le r_k$ be the the distances of the $k$-NN of a query $\query$.
Then, the Maximum Likelihood Estimator for $\query$ at distance $r_k$ is
\begin{equation}
	\hat{LID}_{k}(\query) =
	-\left(
	\frac{1}{k}
	\sum_{i=1}^{k} \ln\frac{r_i}{r_k}
	\right)^{-1}
\end{equation}

\subsubsection{Relative Contrast}

The Relative Contrast (RC) captures the relationship between the
distance of the $k$-th nearest neighbor of a query and the average distance
of points from the query.

\begin{definition}[\cite{DBLP:conf/icml/HeKC12}]
	For a query point $\query$ and a set $S$, let $d_{mean}$ be the average distance of
	$\query$ to points in $S$. The Relative Contrast of $\query$ at $k$ with respect to $S$
	is
	\[
		RC_k(\query) = \frac{d_{mean}}{r_k}
	\]
	where $r_k$ is the distance to the $k$-th nearest neighbor of $\query$ in $S$.
\end{definition}

A small relative contrast implies that the distance of the $k$-th nearest
neighbor is close to the average distance to the query: as a consequence, a hard query
is expected to have a small RC value.

\subsubsection{Query expansion}

The concept of Expansion around a query was first introduced to analyze the
properties of LSH-based indices~\cite{DBLP:conf/soda/AhleAP17}. Here we adopt
the extended definition introduced in~\cite{DBLP:journals/is/AumullerC21}.

\begin{definition}
    Given a query $\query$ and an integer $k$, the Expansion of
	$\query$ at $k$ with respect to $k'>k$ is
    \[
        \Expansion_{k'|k}(\query)=
        \frac{r_{k'}}{r_k}
    \]
\end{definition}

Similarly to the Relative Contrast, a small Expansion around the query
implies that for an index it might be hard to discern between
the $k$-th nearest neighbor and the points that are farther away.
Hence, a hard query is expected to have a small Expansion value.

\subsubsection{$\epsilon$-hardness}

The $\epsilon$-hardness~\cite{DBLP:journals/vldb/ZoumpatianosLIP18} focuses on measuring the number of points that sit in 
the ball of radius $(1+\epsilon)r_1$, where $r_1$ is the distance of the 1st nearest neighbor of the query.
We extend the definition to support the $k$-th nearest neighbor,
and refer to the metric as $\alpha_{\epsilon, k}$ following~\cite{DBLP:journals/vldb/ZoumpatianosLIP18}.

\begin{definition}
For a query point $\query$ and a set $S$, let $r_k$ be the distance of the $k$-th nearest neighbor of $\query$ in $S$. For a parameter $\epsilon > 0$ the $\epsilon$-hardness
is
\[
\alpha_{\epsilon, k}(\query) = \frac{
    \left|
    \left\{x \in S : d(\query, x) \le (1+\epsilon)r_k\right\}
    \right|
}{|S|}
\]
\end{definition}

A high $\epsilon$-hardness value implies that a large fraction of the set $S$ lies at a distance
slightly larger than $r_k$.
Therefore, a high value of the $\epsilon$-hardness is expected for hard queries.

\subsubsection{Empirical hardness}

The aforementioned measures where introduced in the literature to capture different characteristics of the data that possibly relate to the actual hardness of a query.
We now introduce a measure for the work actually invested by an index data structure
to answer a given query. Rather than focusing on the running time, which is implementation and platform dependent, we consider the number
\added{of full distance computations carried out}
by the index while answering the query.
To make the number comparable across different datasets we normalize it by the total dataset size.
Since different index structures are likely to experience different performance on the same query and data, the measures is parameterized by the index structure.

\begin{definition}
    \added{
	Given a data structure $\mathcal{D}$ indexing a point set $S$,
    and a recall threshold $\rho$,
    the empirical hardness $\mathcal{H}_{\mathcal{D},\rho}(\query)$
    of a query $\query$
    is the number of full distance computations carried out by $\mathcal{D}$
    in order to achieve recall $\ge \rho$,
    divided by $|S|$.
    }
\end{definition}

\added{
For instance, consider the index $\mathcal{D}=$HNSW, indexing a
dataset with one million points.
If answering a query $\query$ with recall at least $\rho=0.95$ 
requires computing 100\,000 distances, then the empirical hardness
of $\query$ is $\mathcal{H}_{HNSW,0.95}(\query) = 0.1$.
}

\added{
Note that the definition above considers only \emph{full} distance computations
between points in the original space, as this is usually the most expensive
subroutine by far, while answering a query.
We thus do not count the cost of using of sketches, summaries, or 
similar estimators that are routinely employed to weed out non-relevant
candidate neighbors.
}

Clearly, hard queries correspond to high empirical hardness.

\begin{figure}
    \centering
    \begin{minipage}{.48\columnwidth}
    \includegraphics[width=\textwidth]{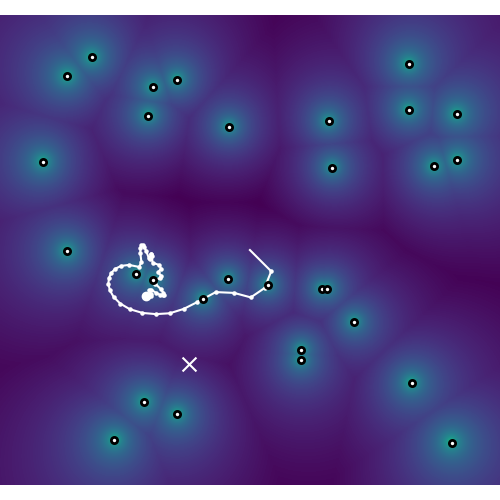}
    \end{minipage}
    \hfill
    \begin{minipage}{.48\columnwidth}
    \includegraphics[width=\textwidth]{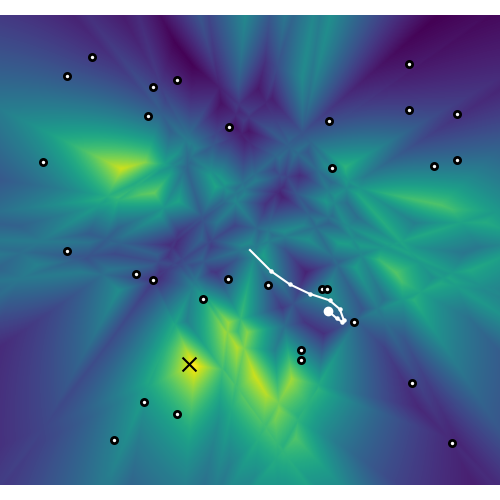}
    \end{minipage}

    \caption{Example dataset of 30 random points in $\mathbb{R}^2$,
    with colors encoding the relative contrast of each position for 
    $k=1$ (left) and $k=10$ (right). High (resp. low) relative contrast is denoted with bright (resp. dark) colors.
    The white paths mark the trajectories followed by \HephGrad to generate a query in each scenario.
    }
    \label{fig:example-rc-landscape}
\end{figure}

\begin{remark}
The hardness of a query, for any of the measures discussed in this section,
is crucially related to its position in the metric space, relative to the other
points in the dataset, \emph{and} to the number of neighbors $k$ we are looking for.
Figure~\ref{fig:example-rc-landscape} reports an example with a dataset of
30 randomly distributed points in $\mathbb{R}^2$.
In the figure, the color encodes the Relative Contrast that a query would have in different positions of the plane, for both $k=1$ and $k=10$.

For $k=1$, notice how the Relative Contrast increases as we get closer to any point:
in fact, getting closer to a point makes it easier to distinguish from the others.
For $k=10$, however, the situation is completely different. The locations of the easiest queries are between the points and not close to them, because the hardness is dictated by how similar the $10$-th nearest neighbor is to the rest of the points.

Remarkably, if we consider a query located at the position marked with the $\times$, we have two very different situations depending on the value of $k$.
For $k=1$ the Relative Contrast of the query is $\approx 1.84$, for $k=10$ its
Relative Contrast is $\approx 4.01$: in the latter case therefore the same query point is comparatively easier.
\end{remark}

\section{Generating workloads}
\label{sec:workloads-generators}

We propose two different ways of generating query workloads:
\HephAnn and \HephGrad\footnote{In Greek mythology, Hephaestus is the god of artisans and blacksmiths.}.
Both algorithms generate queries tailored for a specific dataset $S$ with a query hardness measured according to a scoring function \Scoref: for a point \point and a dataset $S$,
$\Scoref(S, \point)$ is a real number representing how hard or easy the query is.
For instance, \Scoref could compute the Relative Contrast of \point.
The synthesis of each query then aims at reaching a user-supplied range of desired scores $[y_l, y_u]$.
For example, $y_l$ and $y_u$ could be bounds on the desired Relative Contrast of the queries.
In general, we will use as scoring functions the hardness measures defined in the previous section.

We also describe a simpler mechanism to build queries based on adding Gaussian noise to queries, to be used as a baseline in our experimental evaluation.

\subsection{\HephAnn}

\begin{algorithm}[t]
	\caption{\HephAnn\label{alg:heph-ann}}
	\KwIn{
		Dataset $S$;
		starting point $\query$;
		hardness scoring function $\Scoref: (\mathcal{X}, S)\to \mathbb{R}$;
		initial temperature $T$;
		maximum number of iterations $max\_iter$;
		target score range $(y_l, y_h)$.
	}

	\SetKwFor{WithProb}{with probability}{do}{}

	\Let{$y$}{$\Scoref(S, \query)$}\;
	\For{$i \leftarrow 1$ \textbf{to} $max\_iter$}{
		\Let{$\query'$}{move $\query$ randomly to a nearby location}\;
		\Let{$y'$}{$\Scoref(S, \query')$}\;
		\lIf{$y_l \le y' \le y_h$}{
			\Return $\query'$
		}
		\Let{$\Delta$}{$\min\{ |y - y_l|, |y - y_h| \}$}\;
		\Let{$\Delta'$}{$\min\{ |y' - y_l|, |y' - y_h| \}$}\;
		\If(\emph{(the candidate is closer to the target)}){$\Delta' < \Delta$}{
			\Let{$\query$}{$\query'$}\;
			\Let{$y$}{$y'$}\;
		}\Else(\emph{(we accept a worse move with some probability)}){
			\Let{$t$}{$T / i$}\;
			\Let{$p$}{$\exp(-|y - y'|/t)$}\;
			\WithProb{$p$}{
				\Let{$\query$}{$\query'$}\;
				\Let{$y$}{$y'$}\;
			}
		}
	}
    \Return $\query$\;
\end{algorithm}

The first query generation procedure we propose
(Algorithm~\ref{alg:heph-ann}) starts from an arbitrary point \query,
which can be for instance sampled randomly from the
enclosing ball of the dataset.
Then, \HephAnn
iteratively
moves \query so that its score $\Scoref(S, \query)$ 
falls within the target range $[y_l, y_u]$. 

Making no assumption on
$\Scoref$, at every iteration we move $\query$ in a random direction, by a random amount. 
If this
move leads to a position with a hardness score closer to the target, then we
continue to the next iteration. However, the move could of course land into a
position whose score is farther from the target range. In this case, we adopt
the classic \emph{simulated annealing} strategy:
with some probability we
accept the move to a worse position, otherwise we backtrack to the previous
position of \query.
The algorithm terminates when either \query has a score within the required bounds or a maximum number of iterations is reached.

In particular, we adopt the \emph{fast annealing} strategy:
at iteration $i$ the probability of accepting a bad move is $\exp(-|y - y'|/t)$,
where $t=T/i$ is a linearly decreasing \emph{temperature} from an initial value $T$,
and $y=\Scoref(S, \query)$ (resp. $y'=\Scoref(S, \query')$) is the score of \point
(resp. the moved point $\query'$).

The intuition is the following: in the first iterations the algorithm boldly
explores the space, but as it progresses it becomes more and more conservative,
avoiding moves with a worse score.

As for the random distance to move each point in each iteration,
we have to consider different distributions depending on the distance metric employed on each particular dataset.
For datasets using the Euclidean distance candidate queries 
are displaced by a distance sampled from an exponential distribution
$\operatorname{Exp}(\lambda)$. The rate $\lambda$ of the exponential 
distribution is a parameter that we set, by default, to a $100^\textrm{th}$ of the diameter of the dataset. By doing so
we favor small moves while still allowing the occasional long-distance
jump.
For datasets employing the angular distance we apply the same rationale,
with the additional constraint that the angular distance cannot 
exceed 2. Therefore we replace the exponential with 
the beta distribution with parameters $\alpha=0.1$ and $\beta=1$,
whose probability density function is skewed towards 0, thus favoring small tweaks to the position of the point \query.

\subsection{\HephGrad}

\begin{algorithm}[t]
	\caption{\HephGrad\label{alg:heph-grad}}
	\KwIn{
		Dataset $S$;
		starting point $\query$;
		hardness scoring function $\Scoref: (\mathcal{X}, S)\to \mathbb{R}$;
		  learning rate $\eta$;
		maximum number of iterations $max\_iter$;
		target score range $[y_l, y_h]$.
	}

	\For{$i \leftarrow 1$ \textbf{to} $max\_iter$}{
        \Let{$y$}{$\Scoref(S, \query)$}\;
        \lIf{$y_l \le y \le y_h$}{
        \label{ln:heph-grad:stopping}
            \Return $\query$
        }
        \lElseIf{$y < y_l$}{
            \Let{$\query$}{$\query + \eta\nabla\Scoref(S, \query)$}
        }
        \lElse {
            \Let{$\query$}{$\query - \eta\nabla\Scoref(S, \query)$}
        }
	}
    \Return $\query$\;
\end{algorithm}

The approach we discussed above has the advantage of not making any assumption on
the scoring function \Scoref. The price we pay for this flexibility is that the
algorithm explores the space quite aimlessly.

If the scoring function \Scoref is differentiable (as is the case for the
Relative Contrast, for instance) then we can compute its gradient, which gives
the direction of steepest change of score. 
Therefore, for any given candidate query point \query, we can first compute 
$y=\Scoref(S, \query)$, its hardness score with respect to the set of points $S$.
Then we can check if $y\in [y_l, y_u]$, that is if \query satisfies the required hardness constraints.
If this is not the case, then we compute the gradient
$\nabla\Scoref(S, \query)$ and move \query along the gradient direction, by a
step size $\eta$.
Whether the point is moved upwards or downwards along the gradient depends on whether 
its score $\Scoref(S, \query)$ is above or below the target score range.

This approach, deemed \HephGrad, is summarized in Algorithm~\ref{alg:heph-grad}.
Note that, as in the Gradient Descent procedure routinely used for training
neural networks, the step size $\eta$ is not required to be constant across all
iterations.
Furthermore, note that usually in neural network training the Gradient Descent
procedure is used to \emph{minimize} a loss function.
\HephGrad instead strives to find a \query whose score is not necessarily minimal, but rather falls withing a user-given range. As such it is not strictly a gradient \emph{descent} scheme, because the gradient can be followed in either direction. 

Figure~\ref{fig:example-rc-landscape} reports the trajectories followed by the
optimization process of \HephGrad. In the example with $k=1$ (left pane)
the target for the optimization is a query with relative contrast 8 (an easy query),
whereas for $k=10$ (right pane) the algorithm seeks to place a query with relative
contrast 1.05 (thus a difficult query).

\subsection{Targeting the Empirical hardness}

As we shall see in the experimental section, a given fixed value of an explicative hardness measures (e.g. the Relative Contrast) might correspond to a different empirical hardness depending on the dataset and index being used.
In some scenarios it might be desirable to generate queries with a given empirical hardness for a specific index.
For instance, one might be interested in generating a hard query workload for a specific index to investigate the features of the data that force it to spend a lot of effort to provide the answers.

To tackle this scenario, we propose to modify \HephGrad to take as a parameter an
index $\mathcal{D}$ built on the dataset $S$, as well as a range
of admissible empirical hardness values $[h_l, h_h]$.
Then, the stopping condition
is modified to check whether the empirical hardness $\mathcal{H}_{\mathcal{D},\rho}(\query)$ is within the requested range.
If not, then the algorithm uses the gradient \emph{of a differentiable scoring
function} \Scoref to guide the move of \query to the next position.
In fact the empirical hardness $\mathcal{H}_{\mathcal{D},\rho}$ might be, in general,
not differentiable.

As a concrete example, this algorithm can be instantiated
with \messi as the index data structure for computing the empirical hardness,
and the Relative Contrast to guide the moves of \query by means of its gradient.

Note that the empirical hardness might be employed directly with \HephAnn.
As we shall see in the experimental section, however, \HephGrad converges faster.

\section{Experimental Evaluation}\label{sec:experiments}

We aim at answering the following questions:
\begin{itemize}
\item How well do \emph{explicative} hardness measures correlate with the observed \emph{empirical} hardness?
\item What is the empirical hardness of workloads generated by \HephAnn and \HephGrad,
compared with the baselines?
\item Which algorithm between \HephAnn and \HephGrad converges faster?
\item How do the algorithms scale with the size of the dataset?
\item How effective is \HephGrad at generating workloads targeting a given empirical hardness range?
\end{itemize}

\paragraph{Implementation}

We implement all the algorithms using Python 3.12, using JAX~\cite{jax2018github} for automatic differentiation in \HephGrad and Adam~\cite{kingma2017adam} as the optimizer (as implemented in Optax~\cite{deepmind2020jax}).
The learning rate is left as a parameter to be specified in the following.
For the sake of reproducibility, our experimental pipeline is publicly available\footnote{\url{https://github.com/Cecca/workloads-generation/}}.
\added{
We also provide an easy to use standalone Python implementation of our methods\footnote{\url{https://github.com/Cecca/hephaestus/}}.
}

We execute our experimental evaluation on a single machine equipped with a 48-core Intel(R) Xeon(R) CPU E5-2650 v4 processor, clocked at 2.20GHz, with 251GB of RAM and a 3.7 TB SCSI SSD disk.

\paragraph{Datasets}

We consider the following datasets:
\begin{itemize}
	\item
    \dataset{astro} (euclidean)~\cite{astro}: celestial objects represented by 100 million points of dimension 256.
  	\item
    \dataset{deep1b} (euclidean)~\cite{deep1b}: 100 million Deep1B vectors of dimension 96, extracted from the final layers of a convolutional neural network.
	\item
    \dataset{sald} (euclidean)~\cite{sald}: MRI data, containing 100 million points of dimension 128.
 	\item
    \dataset{seismic} (euclidean)~\cite{iris}: recordings from seismic instruments at thousands of stations globally, comprising 100 million 256-dimensional points.
  	\item
    \dataset{glove} (angular)~\cite{pennington2014glove}: 
    1\,192\,512 word embeddings in 100 dimensions derived from 2 billion tweets.
	\item
    \dataset{nytimes} (angular)~\cite{misc_bag_of_words_164}:
    300\,000 New York Times articles embedded in 256 dimensions.
    \item 
    \dataset{text2image} (angular)~\cite{text2image}
    comprises 10 million, 200-dimensional image embeddings as data,
    and text embeddings as queries.
\end{itemize}

\added{
We consider a random sample of 5 million points from each dataset (with the exception of smaller datasets), in order to allow running a large number of
experiments.
Nevertheless, we stress that our method is able to scale to much larger datasets:
Section~\ref{sec:runtime} reports scalability experiments.
}

\paragraph{Baselines}

We consider two baselines for comparing our workload generators.

Each dataset is complemented by a query workload, which we refer to as \textsc{Baseline},
that is routinely used to benchmark index data structures.
This query workload, comprising 10\,000 queries for \dataset{nytimes} and \dataset{glove} and 1\,000 points for the others, is built by extracting points from the dataset itself.

Furthermore, we consider a simple method:
sample
points from the dataset, and perturb their coordinates using Gaussian noise.
In the following we refer to this approach as \Gaussian.

More in detail, to generate a query we 
sample a point $\point$ uniformly at random from $S$,
adding $\mathcal{N}(0, \sigma^2)$ noise on top of each coordinate.
We will use different values for $\sigma$ to control the distance of generated queries from dataset points.

This approach has been used extensively in the literature in the past 30 years to generate synthetic workloads~\cite{Agrawal1993,Faloutsos1994,Rafiei1997,Chan1999,DBLP:conf/kdd/ShiehK08,Camerraisax2p,KashyapK11,Schafer2012,DBLP:journals/pvldb/LinardiP18,DBLP:journals/pvldb/EchihabiZPB18,DBLP:journals/pvldb/EchihabiZPB19,DBLP:conf/kdd/WangP21,DBLP:journals/pacmmod/Wang0WP023}.

\paragraph{Index data structures}

For the sake of generality we consider four different index data structures:
two exact indices and two approximate indices.
As exact indices we consider \messi~\cite{DBLP:conf/icde/PengFP20} and \dstree~\cite{DBLP:journals/pvldb/WangWPWH13}.
As approximate indices we consider \hnsw and \ivf in the implementations
provided by the \texttt{faiss} library~\cite{DBLP:journals/corr/abs-2401-08281}.

\paragraph{Performance Metrics}

First, we assess the relative merits of different hardness measures in terms of
their correlation with the empirical hardness. Our main performance measure to
evaluate generated workloads is thus the empirical hardness as defined in
Section~\ref{sec:hardness-measures}. Here we detail how we compute this hardness
measure for a query \query, dataset $S$ and index data structure $\mathcal{D}$.

Typically, the index $\mathcal{D}$ has several parameters that can be tweaked,
resulting in a different levels of performance, possibly at the expense of the
accuracy (for approximate indices). For instance, \hnsw has parameters
controlling the number of neighbors used in the graph, and the depth of the
exploration at query time.

For a query \query{} and an index $\mathcal{D}$ on a dataset $S$ we perform a
grid search of the parameter space of $\mathcal{D}$. 
For exact indices (i.e. \messi and \dstree), we pick the configuration that attains the smallest number of distance computations.
For approximate indices (i.e. \ivf and \hnsw), we select among all configurations with recall at least $\rho=0.95$ the one with the minimum number of distance computations.
For brevity, in what follows we denote the empirical hardness
as $\mathcal{H}_\mathcal{D}$, omitting the subscript $\rho=0.95$.
Note that this is a particularly stringent requirement, on a query level:
when $k=10$
our setup will select the fastest configuration returning the \emph{exact} result, even for approximate indices.

\added{
It has been observed~\cite{DBLP:journals/is/AumullerC21,VIBE}
that different configurations of the same index data
structure can achieve the same average recall with very different
runtime performances, when applied to batches of queries.
Furthermore, the recall of each query can vary widely~\cite[Fig 10]{DBLP:journals/is/AumullerC21}.
}
\added{
Therefore, in contrast with usual benchmarking setups,
we tune each index on a \emph{per query} basis when computing the
empirical hardness.
}
While expensive to evaluate, this is to ensure that we get an accurate assessment of the empirical hardness for each query.

\subsection{Evaluating hardness Measures}
\label{sec:evaluating-measures}

The aim of our first set of experiments is to capture how well \emph{explicative} hardness measures relate to the empirical hardness.
This is instrumental in deciding which measure is better to use to generate query workloads.
We consider the query sets that are provided with our benchmark datasets. For
each query $\vec{q}$, we consider the empirical hardness $\mathcal{H}_{IVF}(\vec{q})$ for
the IVF index implementation provided by \texttt{faiss}.
Then, we compute the other hardness measures and evaluate the correlation they exhibit
with $\mathcal{H}_{IVF}(\query)$ for each dataset.
Given that the relation is not linear and that we are only interested in comparing how similar
are the \emph{rankings} induced by the hardness measures, we consider the
\emph{Kendall $\tau$} rank correlation coefficient.
As hardness measures we consider $LID_{10}$, $RC_{10}$, $\Expansion_{20|10}$ and $\epsilon$-hardness with $\epsilon\in [0.05, 0.1, 0.5, 1]$.

\begin{table}
\caption{Absolute value of the Kendall rank correlation coefficient between different explicative measures and the $\mathcal{H}_{IVF}$ empirical hardness (best \underline{underlined}, second-best in \textbf{bold}).
\label{tab:correlations}}
\footnotesize
\begin{tabular}{lrrrrrrr}
\toprule
 & astro & deep1b & glove & nytimes & sald & seismic & text2image \\
\midrule
$Exp_{20|10}$ & \textbf{0.69} & \textbf{0.62} & \textbf{0.86} & \textbf{0.70} & \textbf{0.71} & 0.35 & \textbf{0.53} \\
$LID_{10}$ & 0.42 & 0.37 & 0.71 & 0.21 & 0.35 & 0.04 & 0.50 \\
$RC_{10}$ & \underline{0.75} & \underline{0.77} & \underline{0.91} & \underline{0.91} & \underline{0.79} & \underline{0.81} & \underline{0.74} \\
$\alpha_{0.05, 10}$ & 0.27 & 0.08 & 0.59 & 0.19 & 0.13 & 0.23 & 0.33 \\
$\alpha_{0.1, 10}$ & 0.19 & 0.21 & 0.51 & 0.32 & 0.26 & 0.30 & 0.40 \\
$\alpha_{0.5, 10}$ & 0.19 & 0.22 & 0.23 & 0.66 & 0.61 & \textbf{0.52} & 0.45 \\
$\alpha_{1, 10}$ & 0.22 & 0.28 & 0.84 & 0.21 & 0.38 & 0.37 & 0.29 \\
\bottomrule
\end{tabular}
\end{table}

Table~\ref{tab:correlations} reports the absolute value of the Kendall rank
correlation coefficient for each of the hardness measures with
$\mathcal{H}_{IVF}$. 
The measure that correlates most consistently with the empirical hardness is the Relative Contrast.
In some cases the \Expansion and \LID provide comparable results, 
but in general they have a lower correlation with $\mathcal{H}_{IVF}$.
\added{
For the \LID, this is likely due to its sensitivity to noise in its
estimation, especially in regions where nearest neighbors are tightly packed.
For the \Expansion, its lower correlation can be explained by the fact that it is a very local measure, in that it considers only the distances of the $k$-th and $2k$-th neighbors.
In contrast, the \RC considers both local (the $k$-th nearest neighbor) and more global information (the average distance to all points).
}
As for the \ehard, we note that its highest correlation with the empirical hardness
is achieved for different $\epsilon$ values for each dataset, making it hard
to adopt for the purpose of generating workloads.
\added{
Furthermore, the \ehard{} is rather sensitive to the setting of $\epsilon$: small changes to $\epsilon$ can lead to large differences in the measure, depending on the data distribution.
}

Figure~\ref{fig:correlation-scatterplots} 
in Appendix~\ref{sec:detailed-correlation} 
gives a more detailed view of the
relation between the empirical hardness and the \LID, \RC and \Expansion.
\added{Appendix~\ref{sec:other-empirical-hardness} reports similar results for the other indices we consider.}

Based on these results, we select the Relative Contrast as
the hardness measure to be used going forward to guide the workload synthesis.

\subsection{Generating Workloads}
\label{sec:experiment-generating-workloads}

\begin{figure*}
    \includegraphics[width=\textwidth]{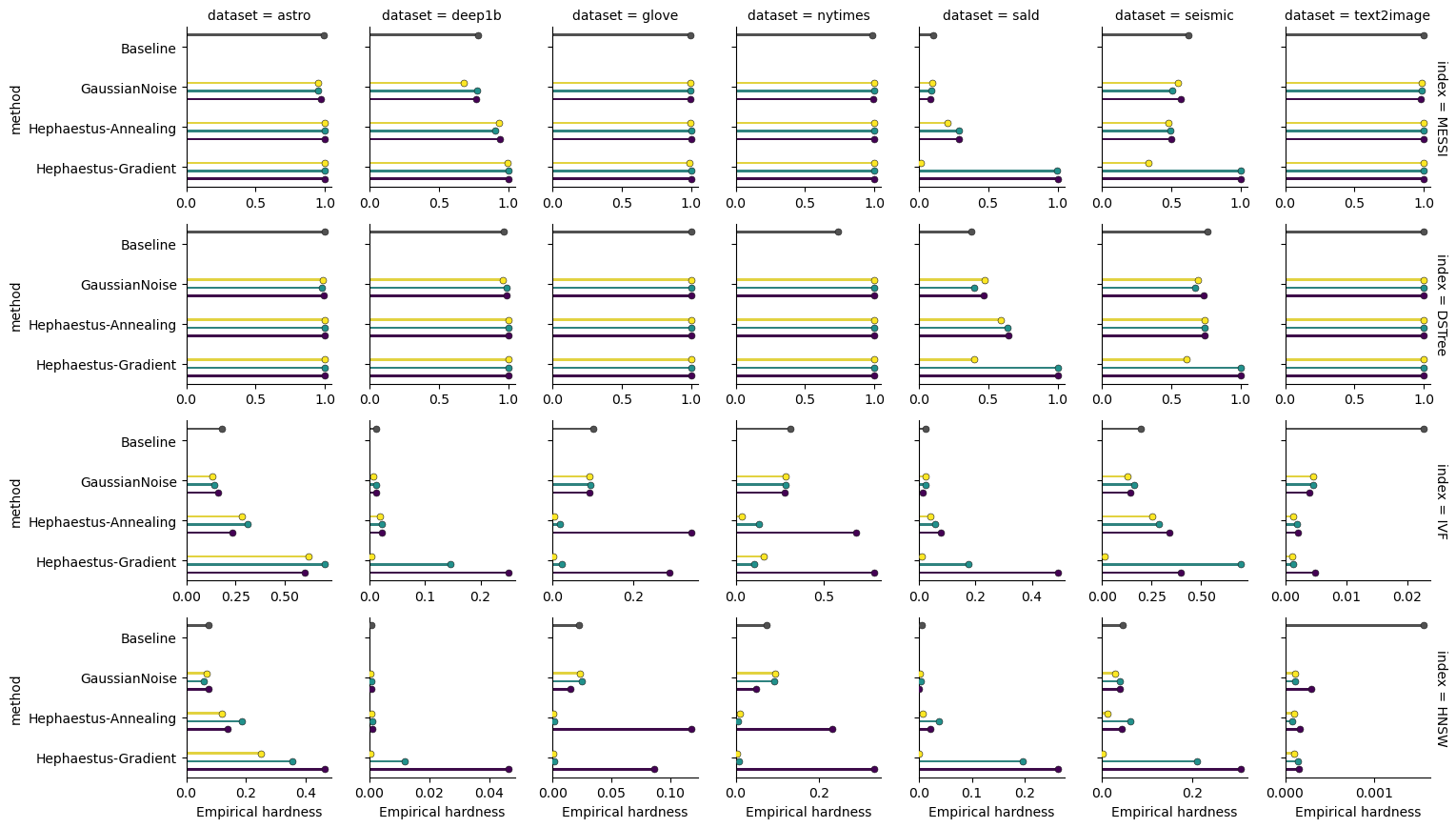}
    \caption[]{Average empirical hardness for different index data structures, over different datasets and
    \ulc{easy}{colorBright},
    \ulc{medium}{colorMedium},
    and \ulc{difficult}{colorDark}
    workloads, for $k=10$.
    \label{fig:index-performance-k10}}
\end{figure*}

In this section we report on experiments using the workload generators described in Section~\ref{sec:workloads-generators}.
Note that we do not consider the method proposed in~\cite{DBLP:journals/vldb/ZoumpatianosLIP18} as the setup is fundamentally different.
In~\cite{DBLP:journals/vldb/ZoumpatianosLIP18}, in fact, for a given fixed query point \query the \emph{dataset} is modified so that \query achieves
the desired hardness level.
In our setting, instead, we keep the dataset fixed and unmodified while generating query points anew.

Using the methods presented in Section~\ref{sec:workloads-generators}
we set out to generate three different workloads for each dataset, deemed \emph{easy}, \emph{medium} and \emph{hard}: the expectation is that index data structures will have to spend more distance computations on the \emph{hard} workload compared to the \emph{easy} one.
We remark that we consider different data structures to ensure that our findings are not specific to a single index: we are \emph{not} aiming at comparing different indices.

For a given dataset $S$ and number of nearest neighbors $k$,
let $\overline{RC}_k$ be the average \RC of the dataset.
The target range of \RC for the generators is
$1+(\overline{RC} -1) \cdot \gamma \pm 5\%$
where $\gamma$ is $0.5$, $0.1$, $0.01$ respectively for
\emph{easy}, \emph{medium} and \emph{hard} queries for Euclidean distance datasets,
and $\gamma=$ $1.5$ (\emph{easy}), $0.5$ (\emph{medium}), $0.25$ (\emph{hard}) for Angular distance datasets.

The difference between the target \RC values between Euclidean and Angular metric spaces is due to the fact that the Angular distance can only take values between 0 and 2, hence very small \RC values are very hard to attain.

For \HephAnn we set the initial temperature to 1 and the maximum number of iterations to 2\,000.
For \HephGrad we set the learning rate to 1 and the maximum number of iterations to 1\,000 (as we shall see, \HephGrad converges faster, hence a smaller maximum number of iterations is reasonable).
Both \HephAnn and \HephGrad, upon reaching the maximum number of iterations, return the last candidate query, whichever its Relative Contrast.
For each hardness level, we generate 30 queries with each generator.

As for the \Gaussian generator, which does not explicitly target the Relative Contrast,
we consider the diameter $\Phi$ of each dataset, setting the standard deviation $\sigma$
of the added noise tto
$\Phi/10^5$,
$\Phi/10^4$, and
$\Phi/10^3$
for \emph{easy}, \emph{medium}, and \emph{hard} queries, respectively.
We generate 100 queries with the \Gaussian generator to account for the larger variability in quality in this queryset.

Figure~\ref{fig:index-performance-k10}
reports the results in terms of average number of distance computations for $k=10$.
Each panel in the figure reports the results on a particular combination
of dataset (arranged in columns) and index data structure (arranged in rows).
Each bar represents the empirical hardness of a workload generated with the method reported on the $y$ axis, with colors encoding the level of expected hardness.
The \texttt{Baseline} entry reports the average number of distance computations on the queryset bundled with each dataset. As such, it is not labelled with any expected hardness, but is included for reference.

First, we observe that the \Gaussian method is not very effective at producing
workloads of different empirical hardness, as this measure is comparable between
\emph{easy} and \emph{hard} workloads on the same dataset/index pair.
In particular, all workloads generated with \Gaussian have an average empirical
hardness comparable with the \textsc{Baseline}.

As for \HephAnn and \HephGrad, if we focus on approximate indices (\ivf and \hnsw),
we observe that \emph{hard} queries are indeed harder that \emph{easy} ones in terms of empirical hardness, in general. This confirms that generating queries with different target ranges of Relative Contrast translates in queries with different empirical hardness.
Furthermore, both methods allow to generate queries that are harder than the \textsc{Baseline} ones, allowing to put more stress on the index data structures.

\begin{figure}
    \centering
    \includegraphics[width=\columnwidth]{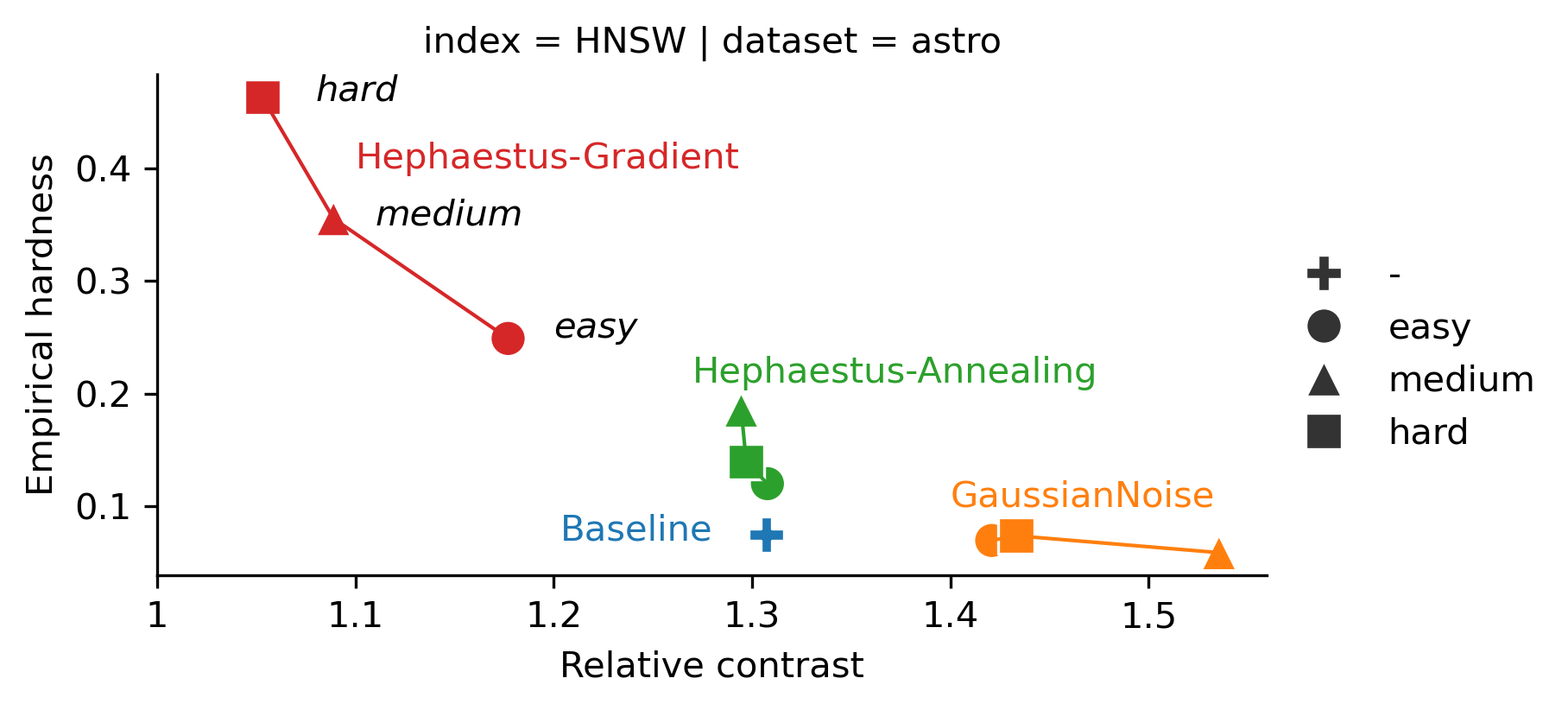}
    \caption{Detailed behavior of the workloads generated by different methods for the \texttt{astro} dataset, evaluated by means of \hnsw. 
    }
    \label{fig:index-focus}
\end{figure}

There are some exceptions: in some cases (e.g. the \ivf index on the
\texttt{astro} dataset), workloads generated with both \HephGrad and \HephAnn
have comparable empirical hardness, irrespective of whether they are supposed to
be \emph{easy} or \emph{hard}.
Another related phenomenon is that in some cases (e.g. \dataset{astro} and index \hnsw),
the empirical difficulty of workloads generated by \HephGrad is higher than \HephAnn, although both methods are set to target the same Relative Contrast ranges.
To investigate why this happens, consider
Figure~\ref{fig:index-focus} that reports, for the \texttt{astro} dataset
and the \hnsw index, the relation between the empirical hardness and the relative
contrast of workloads generated by the different methods, for $k=10$.
This figure corresponds to the data reported in the bottom left panel of Figure~\ref{fig:index-performance-k10}.
Clearly, the relative contrast is higher for the workloads generated by \HephGrad
than the ones generated by \HephAnn, resulting in a lower empirical difficulty.
This is because the generation method did not converge to the target range of RC in the allotted number of iterations.
In fact Figure~\ref{fig:index-focus} shows that all workloads generated by \HephAnn
have the same relative contrast. As we shall see this is due to its slow convergence.

As for exact indices (\messi and \dstree)
in most cases the workloads generated by our methods
have a high empirical hardness.
This stresses the challenge of choosing the right RC range if one wants to generate a query workload with a given empirical hardness for a specific index.
We will tackle this issue, and the ones described in the previous paragraph, in Section~\ref{sec:workload-generation-empirical}.

\subsection{Convergence}\label{sec:convergence}

\begin{figure}
    \includegraphics[width=\columnwidth]{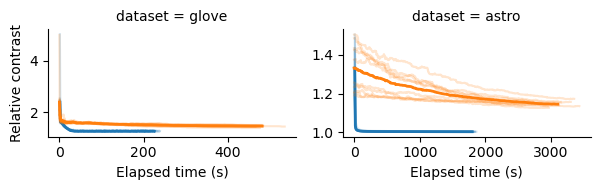}
    \caption[]{
        Convergence of the Relative Contrast against the
        elapsed time for the
        \ulc{\textsc{Hephaestus-Annealing}}{colorOrange} and 
        \ulc{\textsc{Hephaestus-Gradient}}{colorBlue} generators.
        \label{fig:convergence}
    }
\end{figure}

In this section we set out to investigate how fast \HephGrad and \HephAnn
converge.
The setup is as follows:
for each dataset we run 10 instances of \HephAnn and \HephGrad for 1\,000 and
200 iterations, respectively, with each instance generating a single candidate
query.
In each iteration we measure the Relative Contrast of the candidate query and the elapsed time.
In this experiment we set the algorithms to minimize the \RC of generated queries.
Given that the smallest possible \RC is by definition $1$,
this is achieved by setting the target \RC range
to $[1, 1]$ in Algorithms~\ref{alg:heph-ann} and~\ref{alg:heph-grad}
and letting the algorithms run until the maximum number of allowed iterations is reached.

Figure~\ref{fig:convergence} reports, for each dataset, the Relative Contrast of
each of the 10 candidate queries on the $y$ axis, against the elapsed time in
seconds on the $x$ axis using semi-transparent lines.
The solid lines report the average behavior across the 10 instances of each of
the two algorithms.

Clearly, we can observe that \HephGrad converges much faster than \HephAnn to
values of Relative Contrast close to 1. Therefore, if the hardness measure is
differentiable like the Relative Contrast, we recommend using \HephGrad.

Observing the elapsed times, we can notice that different datasets require different times to exhaust the allotted number of iterations. Furthermore, \HephAnn
completes its 1\,000 iterations in about twice the time \HephGrad completes its 200 iterations, suggesting that each iteration of \HephAnn is faster.
We will discuss these aspects in the next section.

\subsection{Running Time and Scalability}
\label{sec:runtime}

\begin{table}[t]
    \caption{
        Running time of \HephAnn (\textsc{H-Ann}) and \HephGrad (\textsc{H-Grad}) to produce workloads
        of \texttt{hard} hardness, both in terms of average time per iteration and overall time.
        \label{tab:times}
    }
    \small
    \begin{tabular}{llrrrr}
\toprule
 &  & \multicolumn{2}{c}{Iteration (s)} & \multicolumn{2}{c}{Total (s)} \\
\cmidrule(r){3-4}
\cmidrule(l){5-6}
 & size & \textsc{H-Ann} & \textsc{H-Grad} & \textsc{H-Ann} & \textsc{H-Grad} \\
\midrule
\multirow[t]{3}{*}{astro} & 5m & 2.3 & 8.1 & 2303.1 & 16.1 \\
 & 10m & 5.7 & 18.5 & 5704.5 & 37.0 \\
 & 15m & 9.3 & 26.4 & 9333.7 & 52.7 \\
\midrule
\multirow[t]{3}{*}{deep1b} & 5m & 1.9 & 5.6 & 1916.5 & 11.1 \\
 & 10m & 4.6 & 14.2 & 4633.3 & 28.5 \\
 & 15m & 7.4 & 19.8 & 7445.0 & 39.6 \\
\midrule
\multirow[t]{3}{*}{sald} & 5m & 2.3 & 6.2 & 2342.7 & 14.3 \\
 & 10m & 4.6 & 13.2 & 4550.7 & 28.9 \\
 & 15m & 7.9 & 21.5 & 7878.5 & 47.4 \\
\midrule
\multirow[t]{3}{*}{seismic} & 5m & 3.2 & 8.2 & 3156.8 & 18.9 \\
 & 10m & 6.1 & 19.9 & 6140.9 & 45.7 \\
 & 15m & 9.0 & 27.6 & 8988.5 & 60.7 \\
\bottomrule
\end{tabular}

\end{table}

We now consider the running time of both \HephAnn and \HephGrad in the
following setup: we generate 10 queries of \emph{hard} hardness under the Relative Contrast measure
and record both the time for each iteration and the overall running time.
To test the scalability, we apply the query generation procedures to samples of the datasets with 5, 10 and 15 million points, omitting from the experiment datasets which have less than 5 million points.
Table~\ref{tab:times} reports the average iteration time and the
overall running time, averaged over the 10 generated queries, in seconds.

Clearly, each iteration of \HephAnn is faster than those \HephGrad, on average. The reason is that the most expensive computation in each iteration of \HephAnn is the computation of the distances from the candidate,
whereas \HephGrad also needs to compute the gradient.

However, as we saw in Figure~\ref{fig:convergence}, \HephGrad
converges much faster.
Hence, the overall running time of \HephGrad is much smaller than that of \HephAnn.

As for the scalability, each method's iteration scales approximately linearly with the dataset size.
This is expected: in both methods the most expensive operation is $O(n)$, with $n$ being the number of points in the dataset.

We omitted from Table~\ref{tab:times} the running time for \Gaussian, for the sake
of conciseness. From a running time perspective, \Gaussian is much faster than
both other approaches, requiring only a few milliseconds to generate each query,
irrespective of the dataset size.
However, as we have seen in the previous sections, the \Gaussian generator offers no control over the hardness of the queries, and typically generates rather easy workloads.

\subsection{Generating Workloads With a Target Empirical hardness}
\label{sec:workload-generation-empirical}

\begin{figure}
    \centering
    \includegraphics[width=\columnwidth]{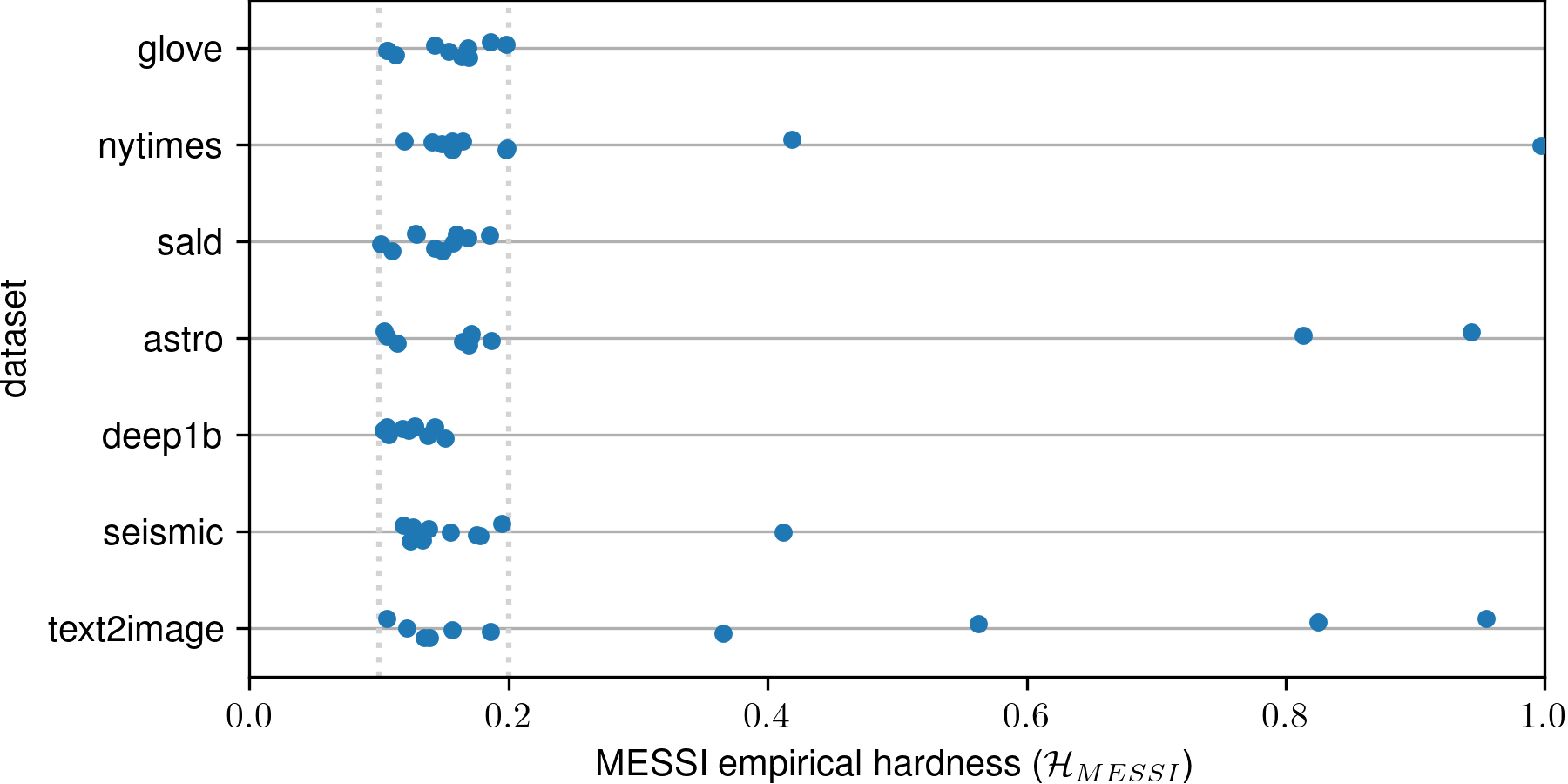}
    \caption{Empirical hardness of a workloads generated with
        \HephGrad, targeting empirical hardness for \messi in 
        the range $[0.1, 0.2]$. Each dot represents one of 
        the 10 generated queries for each dataset.
    }
    \label{fig:empirical-workload}
\end{figure}

In Section~\ref{sec:experiment-generating-workloads}
we observed that most query workloads are rather hard for
exact indices.
It might thus be interesting to generate easy queries in order to investigate the characteristics of queries that exact indices find hard.
To this end, we apply the variant of \HephGrad described in
Algorithm~\ref{alg:heph-grad-empirical} to generate a workload with empirical
hardness between 0.1 and 0.2 for the \messi index.
We generate 10 queries, using the Relative Contrast as the hardness
scoring function $\Scoref$ in Algorithm~\ref{alg:heph-grad-empirical},
and the empirical hardness $\mathcal{H}_{\messi}$ for the stopping condition.
Hence, the Relative Contrast is used to guide the placement of the queries by means of its gradient,
and the empirical hardness is used to assess whether the candidate queries
satisfy the requirements.
We set a maximum of 1\,000 steps.

Figure~\ref{fig:empirical-workload} reports the empirical hardness of the
queries produced by this process, with each poin representing a single query.
Vertical dotted lines mark the target \RC range.
Almost all generated queries exhibit an empirical hardness that is within the
requested bound, on all dataset.
There are couple of exceptions: queries that after the allotted 1\,000
optimization steps still do not fall in the required $\mathcal{H}_{\messi}$
range.
We report them for completeness: in practice such queries can be discarded and
possibly replaced with other queries generated with another run of \HephGrad.

This experiment shows that \HephGrad can be used to target a specific hardness
for a given index, without requiring prior knowledge of the corresponding range
of Relative Contrast values.
In Appendix~\ref{sec:case-study} we will use the queries we just generated to further
investigate the behavior of \messi.

\section{Conclusions}
In this paper, we considered the problem of evaluating and synthesizing
query workloads for benchmarking similarity search approaches, for any kind of similarity search data structure, and for both exact and approximate search.
First, we investigated the relation between \emph{explicative} hardness
measures with the \emph{empirical} hardness encountered by 
index data structures.
We found that the Relative Contrast is the most consistent metric at
characterizing the hardness of a query. Its correlation with the \emph{empirical hardness}
is the highest among the tested measures, but is not perfect.
Devising a new measure that is both more accurate and easy to compute remains an open problem.
We then proposed different methods for generating query workloads. In particular,
we found that to generate workloads whose Relative Contrast falls in a given
range, our method \HephGrad converges rapidly to a solution.
The same method can be used to generate queries that are by construction
hard or easy for a given index.
In future work, we plan to study our method in the context of additional index types and variations~\cite{DBLP:journals/debu/00070P023, DBLP:journals/pvldb/WeiPLP24, DBLP:journals/debu/GaoGXS0WP24, DBLP:journals/pacmmod/AziziEP25, DBLP:journals/pacmmod/WangIP25, DBLP:journals/pacmmod/WeiLLPP25, DBLP:conf/iclr/VallaeysMVD25}.

We note that \HephGrad is independent of the hardness measure, and will benefit from new hardness measures developed in the future.

\begin{acks}
Supported by EU Horizon projects AI4Europe ($101070000$), TwinODIS ($101160009$), ARMADA ($101168951$), DataGEMS ($101188416$), RECITALS ($101168490$), and by $Y \Pi AI \Theta A$ \& NextGenerationEU project HARSH ($Y\Pi 3TA-0560901$).
\end{acks}

\bibliographystyle{ACM-Reference-Format}
\bibliography{workload-generation}

\appendix

\section{Case Study}
\label{sec:case-study}

\begin{figure}
    \centering
    \includegraphics[width=0.49\linewidth]{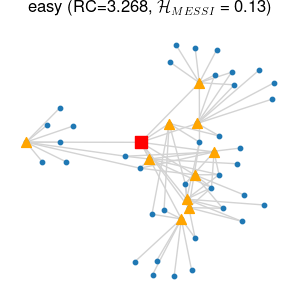}
    \includegraphics[width=0.49\linewidth]{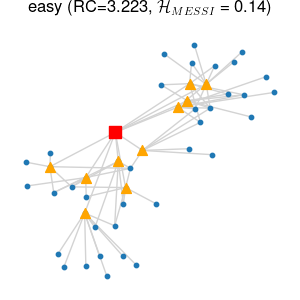}
    \includegraphics[width=0.49\linewidth]{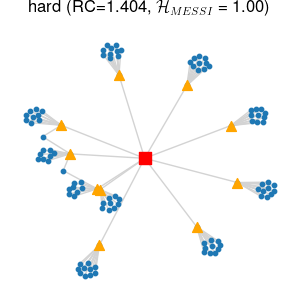}
    \includegraphics[width=0.49\linewidth]{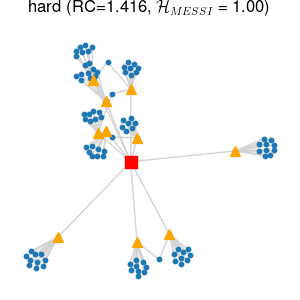}
    \caption{Visualization of the local neighborhood of four different queries 
    on \dataset{glove}, with different empirical hardness related to \messi. 
    The top two queries are easy, while the bottom two queries are hard.
    Red squares represent queries, orange triangles are the 10-nearest neighbors,
    blue circles are the neighbors of the neighbors.
    \label{fig:case-study}}
\end{figure}

To study how the measures and workload generators considered in this paper can 
help in investigating the performance of index data structures, we consider the
\messi index and the \dataset{glove} dataset, for $k=10$.
In particular, we consider four queries generated by \HephGrad. 
The first two queries are easy, from the workload generated in the previous section, with empirical hardness $\mathcal{H}_{MESSI}
\in [0.1, 0.2]$ and Relative Contrast $>3$.
The other two queries are hard, generated from the \textsc{Baseline} workload associated with the dataset,
with $\mathcal{H}_{MESSI} \approx 1$ and Relative Contrast $< 1.5$.

To further characterize the structure of these queries, for each of them
we consider the graph consisting of the
query, its $k$-nearest neighbors (which we deem \emph{immediate} neighbors), and also each neighbor's $k$-nearest neighbors.
In other words, we consider the subgraph of the $k$-nearest neighbor graph comprising
the nodes at hop count at most 2 from the query.
The intuition is that by comparing the immediate neighborhood of the query
with the neighborhood of its $k$-nearest neighbors we can reason about the behavior
of the index.

Figure~\ref{fig:case-study} shows this graph for the four aforementioned queries.
In particular, the red square is the query, and the orange triangles are its $k$-nearest neighbors (i.e. the answer to the query).
The blue dots are the neighbors of each one of the $k$-nearest neighbors of the query.
The graphs are laid out using the \emph{spring} layout from the \texttt{networkx} library~\cite{SciPyProceedings_11}. The length of the drawn edge is influenced by the edge's weight, which we set to the distance between the points the nodes represent.
Therefore, short edges connect nodes corresponding to similar points.
It is worth reminding that this is a 2-dimensional visualization of 100-dimensional points: as such nodes that appear to be close together but \emph{are not} connected by an edge are not, in fact, close in the original space.

The top pair of graphs in Figure~\ref{fig:case-study} represents the two easy queries, whereas the bottom pair is for the two hard queries.
The graphs are remarkably different when comparing hard and easy queries, and similar when comparing queries of the same hardness.

For easy queries, we notice that the nearest neighbors of the query are also mostly nearest neighbors of each other.
On hard queries the opposite is true: the query's neighbors share very
little and their own neighborhoods are quite dissimilar from one another.

Furthermore, on easy queries the immediate neighbors are comparably similar to the query and among them, while being reasonably different from the rest of the dataset, as suggested by the Relative Contrast score above $3.2$.
Conversely, on hard queries the immediate neighbors are closer to their own neighbors than they are to the query, and from the perspective of the query they are similar to the other points of the dataset, as suggested by the Relative Contrast score close to $1.4$.

Consider now that \messi partitions the dataset into a tree data structure
by means of SAX words, i.e. symbolic representations of reduced dimensionality
of the input vectors.
Crucially, these symbolic representations can be used to approximate 
the distance of
the vectors they represent, allowing to restrict the computation of the
distance to fewer candidates.
Considering the examples in Figure~\ref{fig:case-study}, we have that for
easy queries the symbolic representation of the query and its neighbors are
very similar, while being at the same time dissimilar from the others in the
dataset.
For hard queries, though, the approximate distance between symbolic
representations is not accurate enough to allow a meaningful pruning of the
candidates.
An interesting avenue of research is then to allow \messi to better handle
this latter case. A possibility might be to switch to a different space
partitioning scheme for this scenario.

\begin{algorithm}
	\caption{\HephGrad for empirical hardness\label{alg:heph-grad-empirical}}
	\KwIn{
		Dataset $S$;
		starting point $\query$;
		hardness scoring function $\Scoref: (\mathcal{X}, S)\to \mathbb{R}$;
        index data structure $\mathcal{D}$ for $S$;
		  learning rate $\eta$;
		maximum number of iterations $max\_iter$;
		target empirical hardness range $[h_l, h_h]$.
	}

	\For{$i \leftarrow 1$ \textbf{to} $max\_iter$}{
        \Let{$h$}{$\mathcal{H}_\mathcal{D}(\query)$}\;
        \lIf{$h_l \le h \le h_h$}{\label{ln:heph-grad-empirical:stopping}
            \Return $\query$
        }
        \lElseIf{$h < h_l$}{
            \Let{$\query$}{$\query + \eta\nabla\Scoref(S, \query)$}
        }
        \lElse {
            \Let{$\query$}{$\query - \eta\nabla\Scoref(S, \query)$}
        }
	}
    \Return $\query$\;
\end{algorithm}

\section{Additional pseudocode}

We report here the pseudocode (Algorithm~\ref{alg:heph-grad-empirical}) for the 
adaptation of \HephGrad targeting the empirical hardness.

\section{Details on correlations}\label{sec:detailed-correlation}

\begin{figure*}
    \includegraphics[width=\textwidth]{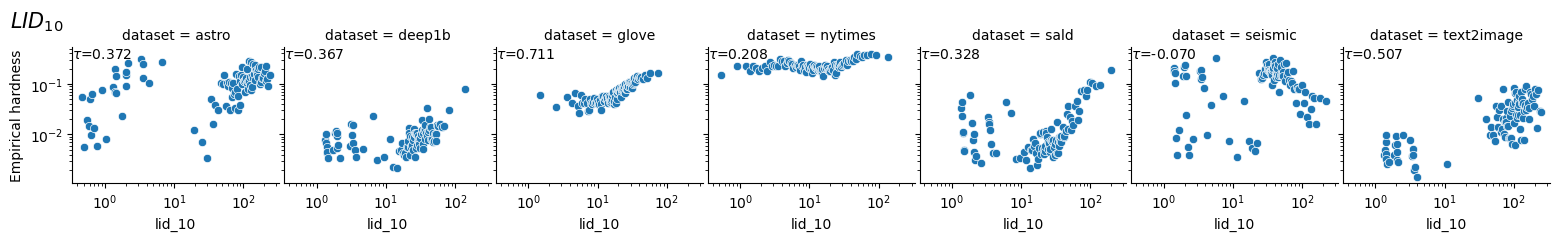}
    \includegraphics[width=\textwidth]{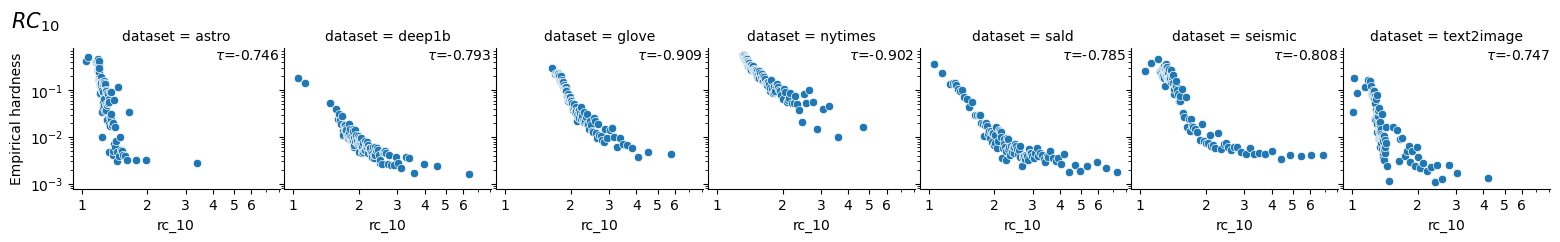}
    \includegraphics[width=\textwidth]{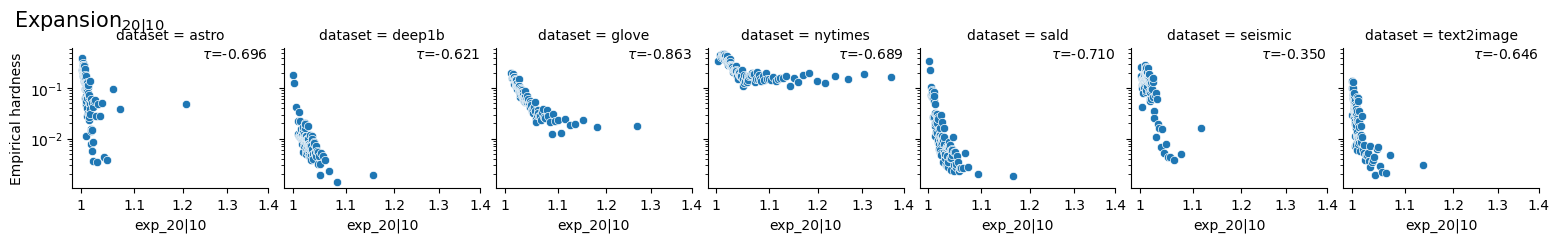}
    \caption{
        Scatterplots of the relation between
        \LID (top row), \RC (mid row), \Expansion{} (bottom row)
        and empirical hardness.
        The $\tau$ reported in each plot is the Kendall rank-correlation coefficient.
        \label{fig:correlation-scatterplots}
    }
\end{figure*}

Figure~\ref{fig:correlation-scatterplots} gives a more detailed view of the
relation between the empirical hardness $\mathcal{H}_{IVF}$ and the \LID, \RC and \Expansion,
which was discussed in Section~\ref{sec:evaluating-measures},
using a logarithmic scale on both axes.
The plots confirm that the Relative Contrast is the measure with the strongest association with the empirical hardness in most cases.
Interestingly, the \Expansion takes values that are very concentrated towards 1, making the association with the empirical hardness very steep.
As for the \LID, an association with the empirical hardness $\mathcal{H}_{IVF}$ can be discerned from the plot, albeit with the presence of many outliers.

\section{Other empirical hardness measures}\label{sec:other-empirical-hardness}

\added{For completeness, Table~\ref{tab:empirical-additional} reports the absolute values of the Kendall rank-correlation
coefficient between the explicative measures and different
empirical hardness  measures, respectively $\mathcal{H}_{HNSW}$,
$\mathcal{H}_{MESSI}$,
and $\mathcal{H}_{DSTree}$.
The highest correlation is underlined, the second-highest is in bold.
We observe that for both \algo{HNSW} and \algo{MESSI} the explicative hardness measure
with the highest correlation is the Relative Contrast. On the other hand,
for \algo{DSTree} the explicative hardness measure with the highest correlation is
the Local Intrinsic Dimensionality.}

\begin{table}
\caption{Absolute value of the Kendall rank correlation coefficient between explicative and empirical hardness (best \underline{underlined}, second-best in bold).\label{tab:empirical-additional}}
\begin{center}
\tiny
\begin{tabular}{llrrrrrrr}
\toprule
&& astro & deep1b & glove & nytimes & sald & seismic & text2image \\
\midrule
\multirow{7}{*}{$\mathcal{H}_{\algo{HNSW}}$}
&$Exp_{20|10}$ & \textbf{0.65} & \textbf{0.68} & \textbf{0.84} & \textbf{0.77} & \underline{0.75} & 0.28 & \textbf{0.57} \\
&$LID_{10}$ & 0.32 & 0.45 & 0.77 & 0.35 & 0.37 & 0.04 & 0.55 \\
&$RC_{10}$ & \underline{0.82} & \underline{0.73} & \underline{0.88} & \underline{0.88} & \textbf{0.72} & \underline{0.79} & \underline{0.67} \\
&$\alpha_{0.05, 10}$ & 0.26 & 0.22 & 0.42 & 0.15 & 0.02 & 0.36 & 0.19 \\
&$\alpha_{0.1, 10}$ & 0.08 & 0.17 & 0.44 & 0.16 & 0.26 & \textbf{0.62} & 0.34 \\
&$\alpha_{0.5, 10}$ & 0.27 & 0.48 & 0.16 & 0.51 & 0.51 & 0.30 & 0.17 \\
&$\alpha_{1, 10}$ & 0.00 & 0.42 & 0.79 & 0.18 & 0.27 & 0.30 & 0.34 \\

\midrule
\multirow{7}{*}{$\mathcal{H}_{\algo{MESSI}}$}
&$Exp_{20|10}$ & \textbf{0.49} & 0.61 & \textbf{0.69} & 0.15 & 0.80 & 0.21 & \underline{0.40} \\
&$LID_{10}$ & 0.38 & 0.32 & \underline{0.69} & \textbf{0.20} & 0.36 & 0.10 & \textbf{0.37} \\
&$RC_{10}$ & \underline{0.69} & \underline{0.92} & 0.65 & 0.07 & \underline{0.92} & \underline{0.94} & 0.36 \\
&$\alpha_{0.05, 10}$ & 0.02 & 0.09 & 0.49 & \underline{0.33} & 0.15 & 0.12 & 0.08 \\
&$\alpha_{0.1, 10}$ & 0.02 & 0.08 & 0.16 & 0.01 & 0.02 & 0.43 & 0.10 \\
&$\alpha_{0.5, 10}$ & 0.25 & 0.53 & 0.44 & 0.08 & \textbf{0.90} & 0.68 & 0.35 \\
&$\alpha_{1, 10}$ & 0.14 & \textbf{0.62} & 0.14 & 0.01 & 0.65 & \textbf{0.74} & 0.25 \\

\midrule
\multirow{7}{*}{$\mathcal{H}_{\algo{DSTree}}$}
&$Exp_{20|10}$ & 0.36 & 0.12 & 0.11 & 0.15 & 0.11 & 0.20 & 0.11 \\
&$LID_{10}$ & \underline{0.52} & \underline{0.48} & 0.01 & 0.17 & \underline{0.39} & \underline{0.56} & 0.16 \\
&$RC_{10}$ & 0.22 & 0.20 & 0.03 & 0.10 & 0.05 & 0.07 & 0.04 \\
&$\alpha_{0.05, 10}$ & 0.32 & 0.34 & \underline{0.59} & \textbf{0.28} & 0.26 & \textbf{0.37} & \underline{0.33} \\
&$\alpha_{0.1, 10}$ & \textbf{0.43} & \textbf{0.47} & 0.52 & \underline{0.33} & \textbf{0.36} & 0.28 & \textbf{0.29} \\
&$\alpha_{0.5, 10}$ & 0.10 & 0.46 & \textbf{0.58} & 0.08 & 0.10 & 0.07 & 0.17 \\
&$\alpha_{1, 10}$ & 0.05 & 0.33 & 0.36 & 0.02 & 0.05 & 0.06 & 0.14 \\
\bottomrule
\end{tabular}

\end{center}
\end{table}

\end{document}